%% file: SHAP_fANOVA.tex
\documentclass[12pt,letterpaper]{article}
\usepackage[cm]{fullpage}
\usepackage[comma, authoryear]{natbib}
\usepackage{array}
\usepackage{multirow}
\usepackage{amsmath, amsthm, amsfonts, amssymb, amscd}
\usepackage{enumerate}
\usepackage[ruled,vlined]{algorithm2e}
\usepackage{hyperref}

\newcommand{\abs}[1]{\lvert #1 \rvert}

\newcommand{\V}{\textmd{Var}}

\newcommand{\E}{\mathbb{E}}

\newcommand{\R}{\mathbb{R}}

\usepackage{xcolor}
\usepackage[normalem]{ulem}

\begin{document}

\title{Statistical Aspects of SHAP: Functional ANOVA for Model Interpretation}
\author{Andrew Herren\thanks{University of Texas at Austin}\\ P. Richard Hahn\thanks{Arizona State University}\\ Rafael Alcantara\footnotemark[1]}
\date{\today}

\maketitle

\begin{abstract}
SHAP (\cite{lundberg2017unified}) is a popular method for measuring variable importance in machine learning models. 
In this paper, we study the algorithm used to estimate SHAP scores and outline its connection to the functional 
ANOVA decomposition. We use this connection to show that challenges in SHAP approximations largely relate to the choice of a feature distribution 
and the number of $2^p$ ANOVA terms estimated. We argue that the connection between machine learning 
explainability and sensitivity analysis is illuminating in this case, but the immediate practical consequences 
are not obvious since the two fields face a different set of constraints. Machine learning explainability 
concerns models which are inexpensive to evaluate but often have hundreds, if not thousands, of features. 
Sensitivity analysis typically deals with models from physics or engineering which may be very time consuming 
to run, but operate on a comparatively small space of inputs.
\end{abstract}

\section{Introduction} \label{intro}
Algorithmic approaches to ``explaining" model predictions have proliferated as machine learning methods gain in popularity. 
We refer interested readers to \cite{molnar2022} or \cite{arrieta2020explainable} for in-depth surveys. 
For the purposes of this paper, we simply note that there are many high-level approaches to explaining 
model predictions and we focus solely on SHAP (\cite{lundberg2017unified}). SHAP is a popular 
``local feature attribution'' method, which means it attempts to explain a model by ``scoring'' 
input feature contributions for a specific prediction. Other common examples of local attribution methods 
include LIME (\cite{ribeiro2016should}), Integrated Gradients (\cite{sundararajan2017axiomatic}), and 
GradCAM (\cite{selvaraju2017grad}). While each of these methods deserve detailed study, this paper is a 
thorough investigation of the statistical properties of SHAP.

SHAP applies the Shapley value from game theory (\cite{shapley1953value}) to model explanation by considering features as 
``players'' in a cooperative game. \cite{lundberg2017unified} approximate Shapley values for each feature using a weighted 
least squares regression, where the regression weights are a transformation of the original Shapley value weights. They refer 
to this method (and the accompanying python library\footnote{\url{https://github.com/slundberg/shap}}) as SHAP, which is the focus 
of this paper. The idea of explaining a model through Shapley values has also appeared several times in earlier 
literature. Both \cite{strumbelj2014explaining} and \cite{datta2016algorithmic} discuss approximations to the Shapley value for 
explaining specific predictions. In the literature on variance-based sensitivity analysis, Shapley values have been used to allocate 
global model variance to specific features (\cite{owen2014sobol}, \cite{song2016shapley}, \cite{owen2017shapley}).

SHAP is popular in industry (\cite{bhatt2020explainable}) and its reach has motivated an active literature 
of debates and proposed improvements. Several papers have presented modifications to SHAP 
that make use of the correlations between features in the training set (\cite{aas2019explaining}, \cite{frye2020shapley}). 
Others have proposed algorithms that augment Shapley value computation with user-specified knowledge of 
causal patterns in the data (\cite{datta2016algorithmic}, \cite{frye2019asymmetric}, \cite{wang2020shapley}). 
\cite{kumar2020problems} argue that many of the above methods have problems that make Shapley values an 
awkward fit for the problem of machine learning interpretability. \cite{kaur2020interpreting} note that many professional 
data scientists misinterpret Shapley values. \cite{chen2020true} respond to many of the concerns noted above, arguing 
that it is up to users to figure out which variety of Shapley value is useful for their problem, but that there is nothing 
wrong with the general approach of using Shapley values. 

This paper seeks to clarify this debate by studying the statistical properties of SHAP, where we note 
connections to the literature on computer experiments and sensitivity analysis. Specifically, SHAP can be represented as 
a into Harsanyi dividends \citep{harsanyi1963simplified} and, under certain assumptions, functional ANOVA effects (\cite{hoeffding1948class}). 
This decomposition enables a formal investigation of two problems that occur frequently in the SHAP literature:
\begin{itemize}
\item How many of the $2^p$ conditional expectations to calculate when approximating Shapley values
\item The choice of a reference, or ``baseline," distribution for each of the conditional expectations
\end{itemize}

While it is well known that there are many possible variants of Shapley value (\cite{sundararajan2020many}), this paper clarifies the technical decisions that a SHAP user must make (or does make implicitly by using the \texttt{shap} library). We hope that this paper helps practitioners move past this debate and proceed cautiously in using SHAP for their specific model.

\section{SHAP overview and notation} \label{overview}

This section introduces both the ``Shapley value" from game theory and the SHAP method in machine learning explainability. 
We show that SHAP attempts to define the Shapley value in the context of model interpretation and then show that this definition 
can be expressed as a linear combination of functional ANOVA components. We will typically use ``Shapley value" to refer 
either to the original game theoretic concept, or to the estimand approximated by SHAP. 

\subsection{Shapley value} \label{shapley-value}

\cite{shapley1953value} considered cooperative games with $n$ players, each of whom can join a coalition with 0 or more other players 
who will receive a collective score. In the economics literature, these scores are often utilities or monetary values, but the mathematics 
of cooperative games only requires that some numeric score be associated with each coalition. Let $\Omega$ refer to the set of $n$ 
players and $2^{\Omega}$ be the set of all possible subsets of $\Omega$. Let $S$ refer to an arbitrary coalition, so that 
$S \in 2^{\Omega}$. The score of a given coalition $S$ is determined by the game's \textit{characteristic function}, $\nu : 2^{\Omega} \rightarrow \R$.
Shapley introduced the following formula, which has come to be known as the ``Shapley value,''
\begin{equation*}
\begin{aligned}
\phi_i\left(\nu\right) &= \sum_{S \subseteq \Omega \setminus \left\{ i \right\}} \frac{\left(\left| S \right| \right)! \left(n - \left| S \right| - 1\right)!}{n!} \left[ \nu\left( S \cup \left\{i\right\} \right) - \nu\left( S \right) \right]
\end{aligned}
\end{equation*}
for player $i$ and characteristic function $\nu$. Broadly speaking, the Shapley value is a weighted average of the contribution 
that player $i$ makes to each of the $2^{n-1}$ coalitions that do not include player $i$. Shapley showed that this formula produces a unique 
score which satisfies the following axioms:
\begin{enumerate}
\item Symmetry: if $\nu \left( S \cup \left\{ i \right\} \right) = \nu \left( S \cup \left\{ j \right\} \right)$ for all $S \in 2^{\Omega}$ and $i \neq j$, then 
$\phi_i\left( \nu \right) = \phi_j\left( \nu \right)$
\item Efficiency: $\sum_{i = 1}^n \phi_i\left( \nu \right) = \nu \left( \Omega \right) - \nu \left( \varnothing \right)$
\item Additivity: For two games with characteristic functions $\nu$ and $\mu$, $\phi_i\left( \nu + \mu \right) = \phi_i\left( \nu \right) + \phi_i\left( \mu \right)$
\end{enumerate}
While it has its origins in theoretical microeconomics, the Shapley value has proven useful in modeling a variety of phenomena. 
We refer interested readers to \cite{roth1988shapley} for a detailed discussion of the broader impact of the Shapley value.

\subsection{SHAP: modified Shapley values for model explainability}

\subsubsection{Definition of players and coalitions} \label{shap-reg-def}

The applicability of Shapley's formula to model explanation hinges on a successful redefinition of the game's players 
and characteristic function in terms of a trained machine learning model. In SHAP, the $p$ features of a model's training set are 
considered ``players'' and the characteristic function is a call to the model's prediction function. While the characteristic function in its 
traditional formulation is a set function, which operates on subsets of players of a game, most model prediction functions require a value 
for every feature that was used in training. At a high level, SHAP's solution to the problem of ``including'' or ``excluding'' features 
from a prediction call is to switch between a ``target value'' for a given feature, which would indicate that the feature was included in a 
coalition, and a ``reference value'' used for non-included features.

To see how this works in more detail, we first introduce some helpful notation and terminology. The \textit{target} is the specific prediction 
that a modeler seeks to explain, and the \textit{baseline} is a ``background'' covariate vector which will replace 
the target value for features that are ``excluded'' from a coalition. Since the construction of coalitions requires switching between 
baseline and target values, we let $z$ refer to a binary vector where 1 indicates use of the target value. Thus, we can map a
coalition $S$ to a synthetic covariate vector, $x_{synthetic}$, as follows:
\begin{equation*}
\begin{aligned}
x_{baseline} &= \left( b_1, b_2, ..., b_p \right)\\
x_{target} &= \left( t_1, t_2, ..., t_p \right)\\
g_i(S) &= \begin{cases} 1 & i \in S\\
0 & i \not\in S
\end{cases}\\
z = g(S) &= \left( g_1(S), ..., g_{p}(S) \right)\\
h(z, x_{baseline}, x_{target}) &= x_{baseline} \times \left( 1 - z \right) + x_{target} \times z\\
x_{synthetic} &= h(g(S), x_{baseline}, x_{target})\\
\end{aligned}
\end{equation*}
where the multiplication terms in the expression $x_{baseline} \times \left( 1 - z \right) + x_{target} \times z$ are both element-wise.

\begin{figure}
\input{figure1.tex}
\caption{Hypercube view of SHAP model evaluation with one baseline value and one target value}
\label{fig:hypercube}
\end{figure}
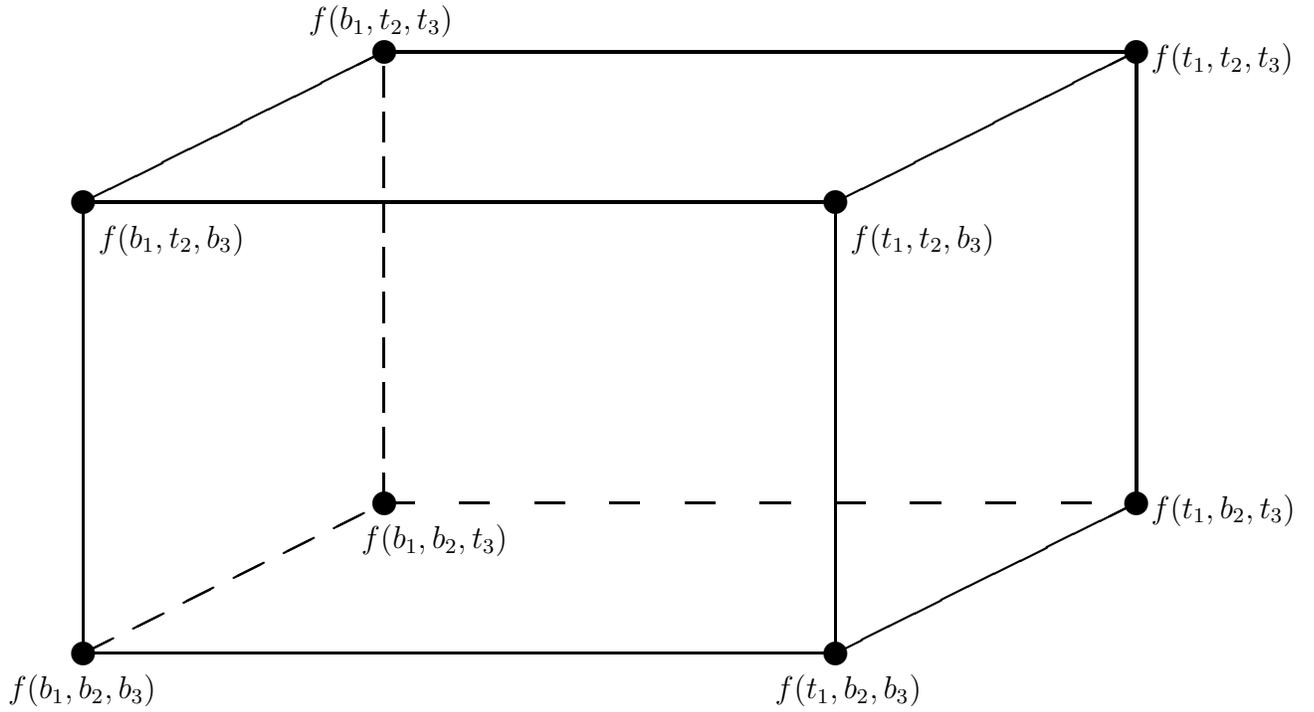

To see this more concretely, suppose a model's training set has 3 real-valued features. 
We define $x_{target} = \left( t_1, t_2, t_3 \right)$ as the vector of covariates corresponding to the \textit{target} and 
$x_{baseline} = \left( b_1, b_2, b_3 \right)$ as the vector of covariates corresponding to the \textit{baseline}. 
Let $f$ be the prediction function for a trained machine learning model. We observe the model predictions for each of the baseline and 
target as $f(b_1, b_2, b_3) = a$ and $f(t_1, t_2, t_3) = b$ and we seek to explain the difference, $b - a$, in terms of each of the three features.
If we consider the three features as ``players,'' then the set $\Omega$ is equal to $\left\{ 1, 2, 3 \right\}$ and we can construct a mapping from each of the 
``coalitions'' to valid vectors in $\R^3$ as follows.
\begin{center}
\begin{tabular}{ |c|c|c| } 
 \hline
 $S$ & $z$ & $x_{synthetic}$ \\ 
 \hline
 $\varnothing$ & $\left( 0, 0, 0 \right)$ & $\left( b_1, b_2, b_3 \right)$ \\ 
 $\left\{ 1 \right\}$ & $\left( 1, 0, 0 \right)$ & $\left( t_1, b_2, b_3 \right)$ \\ 
 $\left\{ 2 \right\}$ & $\left( 0, 1, 0 \right)$ & $\left( b_1, t_2, b_3 \right)$ \\ 
 $\left\{ 3 \right\}$ & $\left( 0, 0, 1 \right)$ & $\left( b_1, b_2, t_3 \right)$ \\ 
 \hline
\end{tabular}
\begin{tabular}{ |c|c|c| } 
 \hline
 $S$ & $z$ & $x_{synthetic}$ \\ 
 \hline
 $\left\{ 1, 2 \right\}$ & $\left( 1, 1, 0 \right)$ & $\left( t_1, t_2, b_3 \right)$ \\ 
 $\left\{ 1, 3 \right\}$ & $\left( 1, 0, 1 \right)$ & $\left( t_1, b_2, t_3 \right)$ \\ 
 $\left\{ 2, 3 \right\}$ & $\left( 0, 1, 1 \right)$ & $\left( b_1, t_2, t_3 \right)$ \\ 
 $\left\{ 1, 2, 3 \right\}$ & $\left( 1, 1, 1 \right)$ & $\left( t_1, t_2, t_3 \right)$ \\ 
 \hline
\end{tabular}
\end{center}
Now we can define the Shapley characteristic function as $\nu\left(S\right) = f(x_{synthetic}) = f\left(h(g(S), x_{baseline}, x_{target})\right)$.
Figure \ref{fig:hypercube} visualizes the synthetic samples created in service of SHAP estimation. 

We can see that the Shapley values for each of the three features are
\begin{equation*}
\begin{aligned}
\phi_1\left(\nu\right) &= \sum_{S \in \Omega \setminus \left\{ i \right\}} \frac{\left(\left| S \right| \right)! \left(3 - \left| S \right| - 1\right)!}{3!} \left[ \nu\left( S \cup \left\{i\right\} \right) - \nu\left( S \right) \right]\\
&= \frac{0! \left(3 - 1\right)!}{3!} \left[ \nu\left( \varnothing \cup \left\{1\right\} \right) - \nu\left( \varnothing \right) \right] + 
\frac{1! \left(3 - 2\right)!}{3!} \left[ \nu\left( \left\{ 2 \right\} \cup \left\{1\right\} \right) - \nu\left( \left\{ 2 \right\} \right) \right] + \\
&\;\;\;\;\; \frac{1! \left(3 - 2\right)!}{3!} \left[ \nu\left( \left\{ 3 \right\} \cup \left\{1\right\} \right) - \nu\left( \left\{ 3 \right\} \right) \right] + 
\frac{2! 0!}{3!} \left[ \nu\left( \left\{ 2, 3 \right\} \cup \left\{1\right\} \right) - \nu\left( \left\{ 2, 3 \right\} \right) \right]\\
&= \frac{1}{3} \left[ f\left( t_1, b_2, b_3 \right) - f\left( b_1, b_2, b_3 \right) \right] + 
\frac{1}{6} \left[ f\left( t_1, t_2, b_3 \right) - f\left( b_1, t_2, b_3 \right) \right] + \\
&\;\;\;\;\; \frac{1}{6} \left[ f\left( t_1, b_2, t_3 \right) - f\left( b_1, b_2, t_3 \right) \right] + 
\frac{1}{3} \left[ f\left( t_1, t_2, t_3 \right) - f\left( b_1, t_2, t_3 \right) \right]\\
\phi_2\left(\nu\right) &= \frac{1}{3} \left[ f\left( b_1, t_2, b_3 \right) - f\left( b_1, b_2, b_3 \right) \right] + 
\frac{1}{6} \left[ f\left( t_1, t_2, b_3 \right) - f\left( t_1, b_2, b_3 \right) \right] + \\
&\;\;\;\;\; \frac{1}{6} \left[ f\left( b_1, t_2, t_3 \right) - f\left( b_1, b_2, t_3 \right) \right] + 
\frac{1}{3} \left[ f\left( t_1, t_2, t_3 \right) - f\left( t_1, b_2, t_3 \right) \right]\\
\phi_3\left(\nu\right) &= \frac{1}{3} \left[ f\left( b_1, b_2, t_3 \right) - f\left( b_1, b_2, b_3 \right) \right] + 
\frac{1}{6} \left[ f\left( t_1, b_2, t_3 \right) - f\left( t_1, b_2, b_3 \right) \right] + \\
&\;\;\;\;\; \frac{1}{6} \left[ f\left( b_1, t_2, t_3 \right) - f\left( b_1, t_2, b_3 \right) \right] + 
\frac{1}{3} \left[ f\left( t_1, t_2, t_3 \right) - f\left( t_1, t_2, b_3 \right) \right]\\
\end{aligned}
\end{equation*}

\subsubsection{Multiple baseline values} \label{mult-baseline}

Section \ref{shap-reg-def} discusses Shapley value estimation in the context of a single baseline value. 
In practice, SHAP users commonly evaluate their target prediction in reference to multiple baselines. We show that notation of the previous 
sections can be extended to cover multiple baselines quite straightforwardly by averaging the Shapley values calculated for each 
individual baseline. 

\begin{table}
\begin{center}
\begin{tabular}{ |c||c|c|c|c||c|c|c|c||c|c| } 
 \hline
 S & $z_1$ & $z_2$ & $z_3$ & $z_4$ & $x_1$ & $x_2$ & $x_3$ & $x_4$ & $4 \choose \abs{S}$ & $\abs{S}$ \\ 
 \hline
 \hline
 $\varnothing$ & 0 & 0 & 0 & 0 & $b_1$ & $b_2$ & $b_3$ & $b_4$ & 1& 0  \\
 \hline
 \hline
 $\left\{1\right\}$ & 1 & 0 & 0 & 0 & $t_1$ & $b_2$ & $b_3$ & $b_4$ & 4 & 1 \\
 $\left\{2\right\}$ & 0 & 1 & 0 & 0 & $b_1$ & $t_2$ & $b_3$ & $b_4$ & 4 & 1 \\
 $\left\{3\right\}$ & 0 & 0 & 1 & 0 & $b_1$ & $b_2$ & $t_3$ & $b_4$ & 4 & 1 \\
 $\left\{4\right\}$ & 0 & 0 & 0 & 1 & $b_1$ & $b_2$ & $b_3$ & $t_4$ & 4 & 1 \\
 \hline
 \hline
 $\left\{1,2\right\}$ & 1 & 1 & 0 & 0 & $t_1$ & $t_2$ & $b_3$ & $b_4$ & 6 & 2 \\
 $\left\{1,3\right\}$ & 1 & 0 & 1 & 0 & $t_1$ & $b_2$ & $t_3$ & $b_4$ & 6 & 2 \\
 $\left\{1,4\right\}$ & 1 & 0 & 0 & 1 & $t_1$ & $b_2$ & $b_3$ & $t_4$ & 6 & 2 \\
 $\left\{2,3\right\}$ & 0 & 1 & 1 & 0 & $b_1$ & $t_2$ & $t_3$ & $b_4$ & 6 & 2 \\
 $\left\{2,4\right\}$ & 0 & 1 & 0 & 1 & $b_1$ & $t_2$ & $b_3$ & $t_4$ & 6 & 2 \\
 $\left\{3,4\right\}$ & 0 & 0 & 1 & 1 & $b_1$ & $b_2$ & $t_3$ & $t_4$ & 6 & 2 \\
 \hline
 \hline
 $\left\{1,2,3\right\}$ & 1 & 1 & 1 & 0 & $t_1$ & $t_2$ & $t_3$ & $b_4$ & 4 & 3 \\
 $\left\{1,2,4\right\}$ & 1 & 1 & 0 & 1 & $t_1$ & $t_2$ & $b_3$ & $t_4$ & 4 & 3 \\
 $\left\{1,3,4\right\}$ & 1 & 0 & 1 & 1 & $t_1$ & $b_2$ & $t_3$ & $t_4$ & 4 & 3 \\
 $\left\{2,3,4\right\}$ & 0 & 1 & 1 & 1 & $b_1$ & $t_2$ & $t_3$ & $t_4$ & 4 & 3 \\
 \hline
 \hline
 $\Omega$ & 1 & 1 & 1 & 1 & $t_1$ & $t_2$ & $t_3$ & $t_4$ & 1 & 4 \\
 \hline
\end{tabular}
\caption{Powerset of coalitions}
\label{tab:samping}
\end{center}
\end{table}

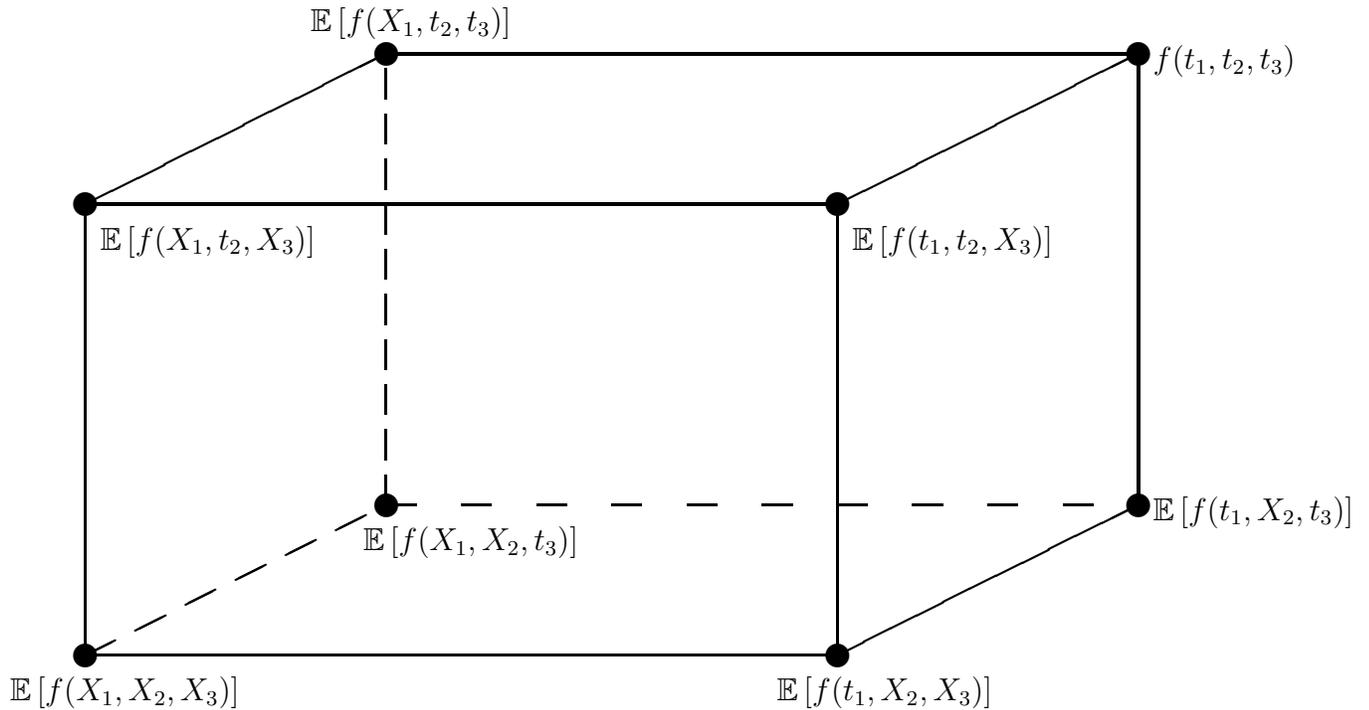
\begin{figure}
\input{figure2.tex}
\caption{Hypercube view of SHAP model evaluation with multiple baselines. Note that all corners of this hypercube except one require computing (or estimating) a conditional expectation of a function over subsets of the random variable $X$.}
\label{fig:hypercube_mult_base}
\end{figure}

Let $Z$ be a $2^p \times p$ binary matrix whose rows are coalitions (1 indicates a feature's inclusion). $Z$ is contructed for $p = 4$ in Table \ref{tab:samping}. 
We let $n$ refer to the number of baseline vectors under evaluation. We refer to the $i$-th baseline as $b^{(i)}$ and the 
target vector as $t$. Let $X^{(i)} = Z t + (1 - Z) b^{(i)}$ be a $2^p \times p$ matrix of ``synthetic'' predictors, determined by the baseline and 
target vectors. Note that for any baseline, $b^{(i)}$, the Shapley values $\phi^{(i)}$ can be estimated by regressing $f(X^{(i)})$ on $Z$ as detailed in 
Section \ref{shap-reg}. Now, we can express the SHAP regression problem with multiple baselines as 
\begin{equation*}
\begin{aligned}
\begin{pmatrix} Z \\ \cdots \\ Z\end{pmatrix} \phi^{*} &= \begin{pmatrix} f(X^{(1)}) \\ \cdots \\ f(X^{(n)})\end{pmatrix}\\
\begin{pmatrix} I & \cdots & I \end{pmatrix} \begin{pmatrix} Z \\ \cdots \\ Z\end{pmatrix} \phi^{*} &= \begin{pmatrix} I & \cdots & I \end{pmatrix} \begin{pmatrix} f(X^{1}) \\ \cdots \\ f(X^{n})\end{pmatrix}\\
n Z \phi^{*} &= f(X^{(1)}) + \cdots + f(X^{(n)})\\
n Z \phi^{*} &= Z \phi^{(1)} + \cdots + Z \phi^{(n)}\\
\phi^{*} &= \frac{\phi^{(1)} + \cdots + \phi^{(n)}}{n}
\end{aligned}
\end{equation*}
Thus, the solution to the SHAP regression problem with multiple baselines is simply the average of the SHAP estimates for each of the individual baselines.
Since expectation is a linear operator, we see that these Shapley values can alternatively be computed as the solution to a regression of the average synthetic 
predictions on $Z$ ($Z \phi^{*} = \frac{1}{n} \left[ f(X^{(1)}) + \cdots + f(X^{(n)})\right]$). If multiple baselines are selected to approximate the sampling distribution 
of $X$, then we can rewrite the hypercube notation of Section \ref{shap-reg-def} as in Figure \ref{fig:hypercube_mult_base}.

\subsubsection{Harsanyi Dividend Decomposition of SHAP and Connection to Functional ANOVA} \label{fANOVA}

Define for each subset $u$ of $\Omega$
\begin{equation*}
\begin{aligned}
f_{u} &= \E\left[f(X) \mid X_u = x_u \right] + \sum_{s = 0}^{\lvert u \rvert - 1}\sum_{v \subset u: \lvert v \rvert = s} \left(-1\right)^{\lvert u \rvert - s} \E\left[f(X) \mid X_v = x_v \right],
\end{aligned}
\end{equation*}
where $X_u$ refers to the variables in $X$ indexed by indices $u$ and $x_u$ refers to a specific realization of $X_u$. $f_u$ is a Harsanyi dividend (\cite{harsanyi1963simplified}) with value function $\nu(u) = \E\left[f(X) \mid X_u = x_u \right]$. Since $\nu(\Omega) = f(x)$, it follows that the function evaluation $f(x)$ can be represented as
\begin{equation*}
\begin{aligned}
f(x) &= \sum_{u \subseteq \left\{1, \dots, p\right\}} f_u
\end{aligned}
\end{equation*}
Independent features also ensure that each $f_u$ in the decomposition above can be interpreted as a functional ANOVA effect.

Functional ANOVA refers to a decomposition of function evaluations into the $2^p$ powerset of orthogonal ``effects'' attributable to subsets of features. Much has been written about the functional ANOVA. We introduce the notation necessary to draw connections to SHAP and refer the interested reader to \cite{hoeffding1948class}, \cite{stone1994use}, \cite{hooker2004discovering}, \cite{hooker2007generalized}, and \cite{liu2006estimating} for more detail.

\cite{hooker2004discovering} define the functional ANOVA recursively in terms of a subset $u \subseteq \left\{1, \dots, p \right\}$ of feature indices, 
\begin{equation*}
\begin{aligned}
f_{u} &= \E\left[f(X) - \sum_{v \subset u} f_v \mid X_u = x_u \right]
\end{aligned}
\end{equation*}
where the variables are assumed to be independent. \cite{hooker2007generalized} generalizes the functional ANOVA to dependent variables, though we lose the simple closed form representation. This paper investigates some sensitivity analysis methods which assume a functional ANOVA interpretation. This does not require the actual underlying features to be independent, but it does necessitate that the conditional expectations in the $f_u$ terms are computed with respect to an independent feature distribution (e.g., by marginalizing over each feature separately).

We now illustrate this decomposition with a specific example. Assume that all features $X_i$ are independently distributed $\textrm{U}\left[0,1\right]$, so that each marginal density 
$p(X_i) = 1$ and that $p = 3$. In this case, $\Omega = \{1,2,3\}$, $X = (X_1, X_2, X_3)$, and the power set of feature combinations is given by 
\begin{equation*}
\begin{aligned}
2^{\Omega} &= \left\{\varnothing,\left\{1\right\},\left\{2\right\},\left\{3\right\},\left\{1,2\right\},\left\{1,3\right\},\left\{2,3\right\},\left\{1,2,3\right\}\right\}
\end{aligned}
\end{equation*}

So the value of a function $f$ evaluated at some realization $x$ of the random variable $X$ can be decomposed as
\begin{equation*}
\begin{aligned}
f(x) &= f_{\varnothing} + f_{1} + f_{2} + f_{3} + f_{12} + f_{13} + f_{23} + f_{123}\\
f_{\varnothing} &= \E\left[f(X_1, X_2, X_3)\right]\\
f_{1} &= \E\left[f(X_1, X_2, X_3) \mid X_1 = x_1 \right] - f_{\varnothing} = \E\left[f(x_1, X_2, X_3) \right] - \E\left[f(X_1, X_2, X_3)\right]\\
f_{2} &= \E\left[f(X_1, X_2, X_3) \mid X_2 = x_2 \right] - f_{\varnothing} = \E\left[f(X_1, x_2, X_3) \right] - \E\left[f(X_1, X_2, X_3)\right]\\
f_{3} &= \E\left[f(X_1, X_2, X_3) \mid X_3 = x_3 \right] - f_{\varnothing} = \E\left[f(X_1, X_2, x_3) \right] - \E\left[f(X_1, X_2, X_3)\right]\\
f_{12} &= \E\left[f(X_1, X_2, X_3) \mid X_1 = x_1, X_2 = x_2 \right] - f_1 - f_2 - f_{\varnothing}\\ 
&= \E\left[f(x_1, x_2, X_3) \right] - \E\left[f(x_1, X_2, X_3)\right] - \E\left[f(X_1, x_2, X_3)\right] + \E\left[f(X_1, X_2, X_3)\right]\\
f_{13} &= \E\left[f(X_1, X_2, X_3) \mid X_1 = x_1, X_3 = x_3 \right] - f_1 - f_3 - f_{\varnothing}\\ 
&= \E\left[f(x_1, X_2, x_3) \right] - \E\left[f(x_1, X_2, X_3)\right] - \E\left[f(X_1, X_2, x_3)\right] + \E\left[f(X_1, X_2, X_3)\right]\\
f_{23} &= \E\left[f(X_1, X_2, X_3) \mid X_2 = x_2, X_3 = x_3 \right] - f_2 - f_3 - f_{\varnothing}\\ 
&= \E\left[f(X_1, x_2, x_3) \right] - \E\left[f(X_1, x_2, X_3)\right] - \E\left[f(X_1, X_2, x_3)\right] + \E\left[f(X_1, X_2, X_3)\right]\\
f_{123} &= \E\left[f(X_1, X_2, X_3) \mid X_1 = x_1, X_2 = x_2, X_3 = x_3 \right] - f_{12} - f_{13} - f_{23} - f_1 - f_2 - f_3 - f_{\varnothing}\\ 
&= f(x_1, x_2, x_3) - \E\left[f(x_1, x_2, X_3)\right] - \E\left[f(x_1, X_2, x_3)\right] - \E\left[f(X_1, x_2, x_3)\right]\\
&\;\;\;\;+ \E\left[f(x_1, X_2, X_3)\right] + \E\left[f(X_1, x_2, X_3)\right] + \E\left[f(X_1, X_2, x_3)\right] - \E\left[f(X_1, X_2, X_3)\right]
\end{aligned}
\end{equation*}

Some arithmetic shows that in this case
\begin{equation*}
\begin{aligned}
\phi_1\left(f\right) &= \frac{1}{3} \left[ \E\left[f(x_1, X_2, X_3)\right] - \E\left[f(X_1, X_2, X_3)\right] \right] + 
\frac{1}{6} \left[ \E\left[f(x_1, x_2, X_3)\right] - \E\left[f(X_1, x_2, X_3)\right] \right] + \\
&\;\;\;\;\; \frac{1}{6} \left[ \E\left[f(x_1, X_2, x_3)\right] - \E\left[f(X_1, X_2, x_3)\right] \right] + 
\frac{1}{3} \left[ f\left( x_1, x_2, x_3 \right) - \E\left[f(X_1, x_2, x_3)\right] \right]\\
&= \frac{1}{3} \left( f_1 \right) + \frac{1}{6} \left( f_{12} + f_1 \right) + \frac{1}{6} \left( f_{13} + f_1 \right) + \frac{1}{3} \left( f_{123} + f_{12} + f_{13} + f_1 \right)\\
&= f_1 + \frac{1}{2} \left( f_{12} + f_{13} \right) + \frac{1}{3} \left( f_{123} \right)\\
\phi_2\left(f\right) &= f_2 + \frac{1}{2} \left( f_{12} + f_{23} \right) + \frac{1}{3} \left( f_{123} \right)\\
\phi_3\left(f\right) &= f_3 + \frac{1}{2} \left( f_{13} + f_{23} \right) + \frac{1}{3} \left( f_{123} \right)\\
\end{aligned}
\end{equation*}

This allows a straightforward interpretation of SHAP as an equal division of functional ANOVA terms for a given ``target'' value $x$, assuming independent features. More broadly, with $p$ features we can write the SHAP estimate for feature $i$ and function $f$ as
\begin{equation*}
\begin{aligned}
\phi_i\left(f\right) &= \sum_{j = 1}^p \frac{1}{j} \sum_{S \subseteq 2^{\Omega}: i \in S, \abs{S} = j} f_S
\end{aligned}
\end{equation*}

Note that this equivalence is not new to the sensitivity analysis literature. \cite{owen2014sobol} decomposes the global Shapley value as above using the variances of the functional ANOVA terms. For examples of references in the context of individual Shapley values, see \cite{keevers2020power}, \cite{hiabu2022unifying}, and \cite{bordt2022shapley}.\footnote{Previous versions of this manuscript did not clarify the difference between the Harsanyi dividend decomposition -- which is always true -- and the functional ANOVA decomposition, which assumes feature independence.}

\subsubsection{Connection to design of experiments} \label{shap-doe}

A central challenge in computing Shapley values in the $p$-dimensional model interpretation setting is that the power set expansion of $2^p$ terms is 
computationally prohibitive for large $p$. This problem is addressed in the model interpretation literature by choosing a small subset of the 
full power set for evaluation. Before discussing the details of this approximation in SHAP (Section \ref{shap-sampling}), we introduce the closely related problem 
of \textit{fractional factorial} design.

\cite{dean2017design} define a \textit{contrast} vector as an $m$-vector of coefficients $c$ such that $\sum_{j = 1}^{m} c_j = 0$ which define an estimator $c y$ 
as a linear combination of $y$ values. In multi-baseline SHAP, the $y$ values correspond to $2^p$ corners ($\E\left[f(X) \mid X_S = x_S\right]$) of the 
hypercube in Figure \ref{fig:hypercube_mult_base} and the contrast coefficients $c$ corresponds to SHAP weights determined by the formula 
$\frac{\abs{S}!\left(p - \abs{S} - 1 \right)!}{p!}$ if $i \not\in S$ and $\frac{\left(\abs{S} - 1\right)!\left(p - \abs{S} - 2 \right)!}{p!}$ otherwise. 

Thus, SHAP estimates for feature $i$ correspond to a contrast in $\E\left[f(X) \mid X_S = x_S\right]$ that estimate 
$\phi_i\left(f\right) = \sum_{j = 1}^p \frac{1}{j} \sum_{S \subseteq 2^{\Omega}: i \in S, \abs{S} = j} f_S$. Traditional experimental design 
literature introduces the notion of a \textit{fractional factorial} design, which economizes the number of experiment runs at the 
expense of estimating some higher order interactions. In the SHAP framing, we can see the utility of this approach in the following example. 
Suppose $f_S = 0$ for all $S$ with $\abs{S} > 1$, $\phi_i\left(f\right)$ can be evaluated for each $i$ as 
$\E\left[f(X) \mid X_i = x_i\right] - \E\left[f(X)\right]$, requiring $2p$ conditional expectation calculations rather than $2^p$. 

Of course, exact knowledge of the nonzero interaction terms is rare. In traditional experiments, fractional factorial designs are often created with 
careful integration of domain knowledge and statistical expertise so that interactions are omitted if prior scientific knowledge suggests factors are 
not related. In the SHAP use case, domain knowledge can also play a role, though it may be difficult in high dimensional problems to identify 
specific interactions that can be excluded. Instead, users may choose sampling plans according to specific hypotheses. One example 
is the hypothesis of \textit{factor sparsity} (\cite{box1986analysis}), which posits that only a small subset of features and their higher-order interactions 
are active. Another example is the hypothesis that higher-order interactions are rare. 

We note briefly that in traditional design of experiments (DoE) problems, the number of factors being studied, $p$, might be relatively manageable, but each sample 
might be time-consuming or costly to collect. In SHAP experiments, each sample collection is simply a call to a machine learning prediction API, which is typically 
efficient on modern computers. The problem with a full factorial design in SHAP is typically in that $2^p$ is an impossibly large number on high-dimensional models. 
Superficially, both approaches involve collecting $n < 2^p$ samples. However, the reasons for doing so are different enough that many approaches which are 
used in DoE (Gaussian process surrogates, \cite{gramacy2020surrogates}, for example) are not always tractable or applicable in explainability. 
In the following section, we show the standard sampling procedure employed by the \texttt{shap} library in approximating Shapley values.

\subsubsection{SHAP sampling} \label{shap-sampling}

We see that the exact Shapley value formula includes $2^{p - 1}$ differences in model conditional expectations, 
for a total of $2^p$ conditional expectations of $f$. This makes Shapley value estimation intractable for large $p$.
To overcome this, \cite{lundberg2017unified} propose a method that approximates these values by deliberately 
sampling coalitions in descending order of $\abs{p / 2 - \abs{S}}$ and approximating the values via weighted linear regression.  
\cite{covert2021improving} discuss some convergence properties of this approximation method and introduce an alternative. 
We will have more to say on SHAP approximations when $p$ is large, but first, we introduce the weighted least squares 
SHAP approximation of \cite{lundberg2017unified}.

First, note using the formula above that the sum of Shapley values for a target $x$ across every feature is 
$\sum_i \phi_i = f(x) - f_{\varnothing}$. Thus, any attempt to estimate $\phi_i$ via linear 
regression introduces the side condition that $\sum_i \phi_i = f(x) - \E\left[f(X)\right]$, so that these two 
estimates cannot be omitted during sampling-based estimation. 
To illustrate the sampling scheme, we now consider a model with $p = 4$ features, given in Table \ref{tab:samping}.

We group coalitions in Table \ref{tab:samping} by size (for example $S_1 = \left\{1\right\}$ and $S_2 = \left\{2\right\}$ are different 
coalitions but $\abs{S_1} = 1 = \abs{S_2}$). Observe that every coalition with $\abs{S} = 1$ has an inverse coalition with with $\abs{S} = 3$ 
($\left\{2,3,4\right\}$ is the inverse of $\left\{1\right\}$, etc...). For $\abs{S} = 2$, on the other hand, there is no inverse coalition. In general, 
with $p$ features, there are $\lfloor (p - 1) / 2 \rfloor$ matching blocks. If $p$ is odd, then every block of coalitions with $\abs{S} = a$ for some $a$ 
will have an inverse block, and if $p$ is even, then there will be 1 ``center block.''

SHAP attempts to enumerate the entire power set, starting with the outermost blocks (1 and 3 in Table \ref{tab:samping}). The sampling process
is iterative and at each step, SHAP determines whether to enumerate an entire block. Suppose a user has specified that 
they would like to draw $m < 2^p$ samples. We defer discussion of the regression weights to the next section, but for now we 
note that the regression weights imply a frequency distribution of samples from each block that is proportional to the 
block's size. During sampling, SHAP uses this implied distribution as a stage gate. SHAP iterates from $i = 1$ to $\lceil (p-1) / 2 \rceil$ 
and at each $i$, SHAP looks at the implied frequency of a block, the number of samples that can be allocated, and determines whether to 
allocate all the samples from block $i$ (and $p - i$ if $i \neq p/2$). Let $k \leq m$ be the number of remaining samples, $j$ be the size of 
block $i$, and $w$ be the target share of samples from block $i$. SHAP will enumerate the entire block if $w \geq \frac{j}{k}$.
Once $w < \frac{j}{k}$, SHAP samples from the remaining blocks uniformly with replacement.

\subsubsection{SHAP regression} \label{shap-reg}

Once a subset of the SHAP coalitions has been sampled, the Shapley values are estimated using weighted linear regression. 
The equivalence between weighted least squares estimation and Shapley values is established in the supplement to \cite{lundberg2017unified}
\footnote{The supplemental files can be accessed at \url{https://papers.nips.cc/paper/2017/hash/8a20a8621978632d76c43dfd28b67767-Abstract.html}}, 
but we derive it below using a slightly different approach which will make clear the relation to experimental design. 
At a high level, the goal is to fit a regression model to the synthetic data whose coefficients are the exact Shapley values 
when the full powerset of coalitions is observed and will approximate the Shapley values when the data are a subsample of coalitions.

Let $Z$ be a binary matrix whose rows are coalitions (1 indicates a feature's inclusion). Assume for now that $Z$ is a 
$(2^p - 2) \times p$ matrix, so that all of the coalitions with $0 < \abs{S} < p$ have been sampled. The weighting function \cite{lundberg2017unified} 
use for the SHAP regression is 
\begin{equation*}
\begin{aligned}
w_i(Z) &= \frac{p-1}{{p \choose s} s \left( p - s\right)} = \frac{\left(p-1\right)\left(p - s - 1\right)! \left(s - 1\right)!}{p!}\\
s &= \sum_{j=1}^p Z_{ij}
\end{aligned}
\end{equation*}
This function is undefined when $Z_{i\cdot}$ is the one or zero vector (corresponding to $f(x)$ and $\E\left[f(X)\right]$, respectively), 
which is why the regression matrix $Z$ only has $2^p - 2$ rows. However, we still incorporate these two vectors in estimation via the side condition expressed above. 

Let $y_t = f(x)$ and $y_b = \E\left[f(X)\right]$ and we express the regression model below. First, we define the vector of conditional 
expectations corresponding to hypercube corners in Figure \ref{fig:hypercube_mult_base}. For each row $Z_i$ in the $Z$ matrix, let $S_i$ 
refer to the columns $j$ for which $Z_{ij} = 1$. Now define $y_i = \E\left[f(X) \mid X_{S_i} = x_{S_i}\right]$.
Letting $j$ be a $p$-dimensional vector of ones, we can approximate $y$ via a linear model with side conditions.
\begin{equation*}
\begin{aligned}
y &= Z\beta + \varepsilon\\
j'\beta &= y_t - y_b\\
\end{aligned}
\end{equation*}
We also define a matrix of weights as 
\begin{equation*}
\begin{aligned}
W &= \begin{pmatrix} 
w_1 & 0 & 0 & ... & 0 \\
0 & w_2 & 0 & ... & 0 \\
0 & 0 & w_3 & ... & 0 \\
... & ... & ... & ... & ... \\
0 & 0 & 0 & ... & w_{(2^p-2)}
\end{pmatrix}
\end{aligned}
\end{equation*}

We estimate the Shapley values by fitting a constrained weighted least squares regression using the 
synthetic $Z$ matrix, the predictions $y$, the weight matrix $W$, $f(x)$ and $f_{\varnothing}$. 
The derivation of the regression solution and its equivalence to Shapley values is given in detail in 
Appendix \ref{shap-reg-deriv}. Defining $I$ as the $p$-dimension identity matrix and $J$ as a $p \times p$ matrix of all ones, 
we note that the regression solution is given by
\begin{equation*}
\begin{aligned}
\hat{\beta} &= \left( \frac{p}{p-1} I - \frac{1}{p - 1} J \right) Z'W y + \frac{j\left(y_t - y_b\right)}{p}
\end{aligned}
\end{equation*}
which corresponds to a weighted contrast estimate of the $2^p$ predictions. It is worth noting that the 
derivations in Appendix \ref{shap-reg-deriv} simply demonstrate that the SHAP regression estimator 
returns Shapley values when the entire power set of coalitions is available. \cite{lundberg2017unified} 
show through a simulation study that the SHAP regression estimates can converge to the exact 
Shapley values with $m << 2^p$ samples. \cite{covert2021improving} study the convergence to exact 
Shapley values analytically.

\section{Estimation decisions} \label{approx}

It was shown above that SHAP decomposes into Harsanyi dividends which can be interpreted as functional ANOVA terms in certain cases. Since there are $2^p$ Harsanyi dividends, when $p$ is large, users cannot typically compute $2^p$ conditional expectations. Thus, users must select some subset of the full power set of feature interactions. This is the first significant choice a user must make in estimating Shapley values for their model, although most users make this decision implicitly by using the \texttt{shap} library which samples as in Section \ref{shap-sampling}.

In addition to the sampling challenges outlined above, there is also the question of how to approximate the requisite conditional expectations. In real world scenarios, it would be highly unusual to have access to an analytical formula for the joint distribution, $p(X)$, of the features. Thus, even if a model provides parameters that can be converted into an analytical formula $y = f(X)$, analytically computing the expectation of $f(X) \mid X_{u}$ for any subset $u$ of feature interactions is impossible without further assumptions. $p(X)$ must be estimated from the data or assumed directly by the user, introducing the second significant user decision.

Consider approximating an expected value with a sample $X$ of size $n$ and evaluating $S \subseteq 2^{\Omega}$ of the Harsanyi dividends. These two decisions thus correspond to the choice of $S$ and then choice of $X$.

\subsection{Selecting interaction terms to evaluate} \label{sampling-interactions}

The default sampling scheme for conditional expectations implemented in the \texttt{shap} library is the paired sampling approach 
described in Section \ref{shap-sampling}.

\subsubsection{The current implementation of sampling in the \texttt{shap} library} \label{paired-sampling}

The \texttt{shap} sampling procedure first draws subsets $s$ with $\abs{s} = 1$ and $\abs{s} = p - 1$ and proceeds ``inwards" in decreasing 
order of $\abs{p/2 - \abs{s}}$. We show below that this procedure is generally effective if lower-order interactions dominate, because the 
first and second order interactions are properly aliased (according to the share of interactions described in Section \ref{fANOVA}) after 
only $2p$ samples are taken. 

We demonstrate this empirically for one example. Consider the case of $p = 6$ in which we sample a $12 \times 6$ matrix of subsets 
$s$ with $\abs{s} = 1$ and $\abs{s} = 5$
\begin{equation*}
\begin{aligned}
X_{r} &= \begin{pmatrix}
1 & 0 & 0 & 0 & 0 & 0 \\
0 & 1 & 0 & 0 & 0 & 0 \\
0 & 0 & 1 & 0 & 0 & 0 \\
0 & 0 & 0 & 1 & 0 & 0 \\
0 & 0 & 0 & 0 & 1 & 0 \\
0 & 0 & 0 & 0 & 0 & 1 \\
1 & 1 & 1 & 1 & 1 & 0 \\
1 & 1 & 1 & 1 & 0 & 1 \\
1 & 1 & 1 & 0 & 1 & 1 \\
1 & 1 & 0 & 1 & 1 & 1 \\
1 & 0 & 1 & 1 & 1 & 1 \\
0 & 1 & 1 & 1 & 1 & 1
\end{pmatrix}
\end{aligned}
\end{equation*}
The matrix of second order interaction terms can be split into 5 sub-matrices 
indexed as $X_{ji}$ with $j$ as the leading interaction variable.
\begin{equation*}
\begin{aligned}
X_{1i} &= \begin{pmatrix}
0 & 0 & 0 & 0 & 0\\
0 & 0 & 0 & 0 & 0\\
0 & 0 & 0 & 0 & 0\\
0 & 0 & 0 & 0 & 0\\
0 & 0 & 0 & 0 & 0\\
0 & 0 & 0 & 0 & 0\\
1 & 1 & 1 & 1 & 0\\
1 & 1 & 1 & 0 & 1\\
1 & 1 & 0 & 1 & 1\\
1 & 0 & 1 & 1 & 1\\
0 & 1 & 1 & 1 & 1\\
0 & 0 & 0 & 0 & 0
\end{pmatrix}
X_{2i} &= \begin{pmatrix}
0 & 0 & 0 & 0\\
0 & 0 & 0 & 0\\
0 & 0 & 0 & 0\\
0 & 0 & 0 & 0\\
0 & 0 & 0 & 0\\
0 & 0 & 0 & 0\\
1 & 1 & 1 & 0\\
1 & 1 & 0 & 1\\
1 & 0 & 1 & 1\\
0 & 1 & 1 & 1\\
0 & 0 & 0 & 0\\
1 & 1 & 1 & 1
\end{pmatrix}
\end{aligned}
\end{equation*}

\begin{equation*}
\begin{aligned}
X_{3i} &= \begin{pmatrix}
0 & 0 & 0\\
0 & 0 & 0\\
0 & 0 & 0\\
0 & 0 & 0\\
0 & 0 & 0\\
0 & 0 & 0\\
1 & 1 & 0\\
1 & 0 & 1\\
0 & 1 & 1\\
0 & 0 & 0\\
1 & 1 & 1\\
1 & 1 & 1
\end{pmatrix}
X_{4i} &= \begin{pmatrix}
0 & 0 \\
0 & 0 \\
0 & 0 \\
0 & 0 \\
0 & 0 \\
0 & 0 \\
1 & 0 \\
0 & 1 \\
0 & 0 \\
1 & 1 \\
1 & 1 \\
1 & 1
\end{pmatrix}
X_{5i} &= \begin{pmatrix}
0 \\
0 \\
0 \\
0 \\
0 \\
0 \\
0 \\
0 \\
1 \\
1 \\
1 \\
1
\end{pmatrix}
\end{aligned}
\end{equation*}

The weight terms for the regression matrix $X_r$ are all $\frac{p-1}{{p \choose \abs{s}} \abs{s} (p - \abs{s})} = 1/6$. 
One way to impose the side condition that $\sum_i \phi_i = f(x_t) - f(x_b)$, is to subtract one of the $X_r$ columns from the 
others. Performing this operation defines a new $2p \times p - 1$ main effect matrix which we denote $X_r^{*}$ and a new 
set of interaction matrices $X_{ji}^{*}$ with the $p$-th column of $X_r$ subtracted. Observe that

\begin{equation*}
\begin{aligned}
X_r^{*'} W X_r^{*} &= (1/3)I + (1/3)J\\
\left( X_r^{*'} W X_r^{*} \right)^{-1} &= 3I - (1/2)J\\
X_r^{*'} W X_{1i}^{*} &= \begin{pmatrix}
1/2 & 1/2 & 1/2 & 1/2 & 1/3\\
1/2 & 1/3 & 1/3 & 1/3 & 1/6\\
1/3 & 1/2 & 1/3 & 1/3 & 1/6\\
1/3 & 1/3 & 1/2 & 1/3 & 1/6\\
1/3 & 1/3 & 1/3 & 1/2 & 1/6
\end{pmatrix}\\
X_r^{*'} W X_{2i}^{*} &= \begin{pmatrix}
1/3 & 1/3 & 1/3 & 1/6\\
1/2 & 1/2 & 1/2 & 1/3\\
1/2 & 1/3 & 1/3 & 1/6\\
1/3 & 1/2 & 1/3 & 1/6\\
1/3 & 1/3 & 1/2 & 1/6
\end{pmatrix}
\end{aligned}
\end{equation*}

And thus, for these first two sets of interactions, the alias matrix is given by 
\begin{equation*}
\begin{aligned}
\left( X_r^{*'} W X_r^{*} \right)^{-1} X_r^{*'} W X_{1i}^{*} &= \begin{pmatrix}
1/2 & 1/2 & 1/2 & 1/2 & 1/2\\
1/2 & 0 & 0 & 0 & 0\\
0 & 1/2 & 0 & 0 & 0\\
0 & 0 & 1/2 & 0 & 0\\
0 & 0 & 0 & 1/2 & 0
\end{pmatrix}\\
\left( X_r^{*'} W X_r^{*} \right)^{-1} X_r^{*'} W X_{2i}^{*} &= \begin{pmatrix}
0 & 0 & 0 & 0\\
1/2 & 1/2 & 1/2 & 1/2\\
1/2 & 0 & 0 & 0\\
0 & 1/2 & 0 & 0\\
0 & 0 & 1/2 & 0
\end{pmatrix}
\end{aligned}
\end{equation*}
which is exactly the alias pattern given by the full design matrix with $2^p - 2$ samples. 
Thus, with only $2p$ samples, Shapley regression estimates will be ``exact" in the sense of properly allocating interactions 
if the model consists of first and second order interactions. In this sense, the \texttt{shap} sampling strategy can be an effective 
way to sample from the $2^p$ possible rows of the design matrix.


\subsubsection{An alternative sampling approach: numeric interaction testing} \label{numeric-search}

While the \texttt{shap} sampling approach appears to work well upon empirical investigation (and is explored in theoretical terms in \cite{covert2021improving}), 
the question may remain: to what extent is the model dominated by low-order interaction terms?
From the derivations in Section \ref{overview}, we can see that if $f_S = 0$ for all $S$ with $\abs{S} > 1$, 
then $\phi_i = f_i$ and $\sum_i f_i = f(x) - f_{\varnothing}$. Of course, since $f_S \in \R$, the converse is not true. 
\cite{hooker2004discovering} defines $\sigma^2_S(f_S) = \E\left[f_S^2\right]$ where the expectation is taken with respect 
to features $X_S$ which were conditioned in $f_S$. We can see that if $\sigma^2_S(f_S) = 0$ and $\E\left[f_S\right] = 0$ (as is commonly established in the 
functional ANOVA decomposition), then $f_S = 0$ for all $x$. Throughout Section \ref{sampling-interactions}, we assume that the functional ANOVA interpretation holds, which is to say that the conditional expectations in each $f_u$ are computed with respect to a product measure feature distribution.

One approach to numerically screening interactions terms introduced in \cite{hooker2004discovering} is to search for the smallest 
set of functional ANOVA terms $\mathcal{S}$ such that $\sum_{s \in \mathcal{S}} \sigma_s^2(f_s)$ accounts for a 
pre-specified share of the variance of $f(X)$. \cite{hooker2004discovering} introduces two different algorithms representing 
different inductive biases: a ``breadth-first" algorithm which assumes higher order interaction effects are more likely to be null and a ``depth-first'' 
algorithm which assumes factor sparsity. We present a modified version of the breadth-first algorithm below and show how it can be 
used in conjunction with SHAP and a suitable interaction scoring criteria $\psi(S)$. Note that the substance of the algorithm below 
is as introduced in \cite{hooker2004discovering}, but the presentation and notation is updated here to reflect the intended 
application and previously-introduced notation. Let $\varepsilon \in [0,1]$ be specified by the user.

\begin{algorithm}[H]
\SetAlgoLined
\KwResult{List of functional ANOVA terms to include in SHAP approximation}
$V_t = \V(f(X))$\;
$V_s = 0$\;
$U$ = $\{\}$\;
\For{j = 1 to p}{
  $\mathcal{S} = \{S \subseteq 2^{\Omega}: \abs{S} = j\}$\;
  $\Psi$ = $\{\psi(s)$ for $s$ in $\mathcal{S}\}$\;
  $I$ = argsort($\Psi$)\;
  \For{k = 1 to ${p \choose j}$}{
    $s = \mathcal{S}_{I_k}$\;
    $U$ = $U \cup s$\;
    $V_s = V_s + \hat{\sigma}_s(f_s)$\;
  \eIf{$V_s > V_t(1 - \varepsilon)$}{
   break\;
   }{
   continue\;
  }
 }
}
\caption{Breadth-first search for low-order functional ANOVA terms}
\label{alg:bfs}
\end{algorithm}

Since the procedure above involves the global $\hat{\sigma}_s(f_s)$, it can be used to identify a sufficient set of 
features for computing Shapley values on any target value for a given model. In particular, it provides a approximate 
empirical measure of ``convergence'' since $V_s / V_t > 1 - \varepsilon$. This is of course only helpful and desirable 
if the resulting $U$ is a considerably reduced subset of $2^{\Omega}$. However, if the algorithm fails to stop at a small 
subset of $2^{\Omega}$, that can also provide the user with valuable information about the quality of their Shapley value 
estimates with less than $2^p$ conditional expectations.
\cite{hooker2004discovering} also notes that when $p$ is large, this algorithm can be configured to stop at a given interaction order, 
thus providing a tractable way of signaling that higher order interactions may be present and the user should take care in 
interpreting Shapley values. Variants of this algorithm are appealing in that they enable a numeric search that can be tailored to the user's inductive bias. 
For example, \cite{hooker2004discovering} provides an depth-first algorithm which corresponds to the factor sparsity hypothesis. 
With the computational infrastructure in place to compute $G_s(x)$ for subsets $s$ of $2^{\Omega}$, researchers can modify the search algorithm to prioritize a hypothesized feature structure.

If the algorithm has been run successfully and a subset $U$ of $2^{\Omega}$ has been identified, we can reduce the 
computational burden of SHAP estimation in either of two ways. First, noting that $\phi_i\left(f\right) = \sum_{i = 1}^p \frac{1}{i} \sum_{S \subseteq \Omega: \abs{S} = i} f_S$, 
we can simply compute $f_S$ for every $S \in U$ and combine their estimates according to the formula for $\phi_i\left(f\right)$. 
Second, we can construct a regression matrix $Z$ whose rows correspond only to the coalitions $S \in U$ and 
estimate $\phi_i$ via weighted least squares. 

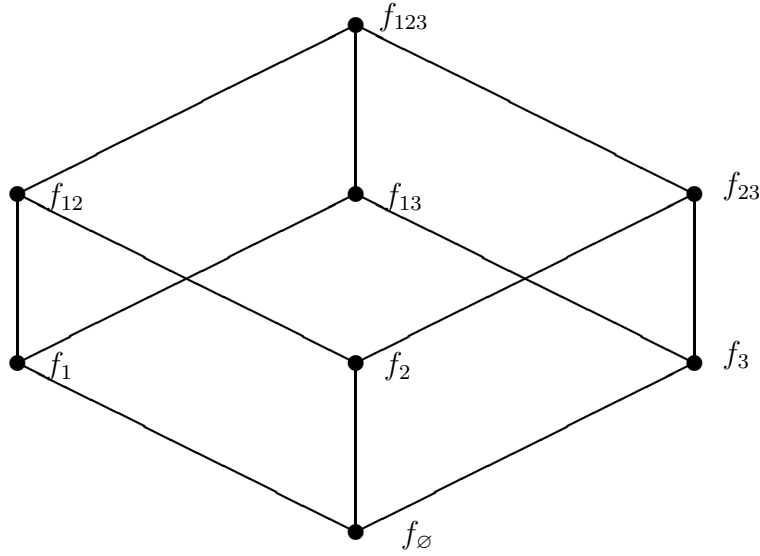
\begin{figure}
\begin{centering}
\input{figure3.tex}
\caption{Hasse diagram of connections between functional ANOVA terms. Excluding an interaction or main effect from the evaluation set involves pruning the estimable functional ANOVA terms in the above diagram. When a small subset of this lattice explains most of the variability of $f(X)$, Shapley values may be approximated well with less computational burden.}
\label{fig:lattice_shap}
\end{centering}
\end{figure}

Now, we discuss one approach to estimating an interaction importance measure $\psi(S)$.
For $p = 3$, we can visualize the SHAP terms as a power set, organized in rows according to $\abs{s}; s \subseteq 2^{\Omega}$ as 
in Figure \ref{fig:lattice_shap}.
\newpage

Now, consider the effect of removing $s = \{1\}$ from the lattice above, which leaves a new structure 
defined in Figure \ref{fig:lattice_shap_1} below. Thus, the effect of removing $s = \{1\}$ is that $f(x)$ must be 
approximated as $f_{\varnothing} + f_2 + f_3 + f_{23} = \E\left[f(X_1, x_2, x_3)\right]$
\begin{figure}
\begin{centering}
\input{figure4.tex}
\caption{Pruned Hasse diagram of functional ANOVA terms with feature 1 removed. Approximating the Shapley value with these terms requires $4 < 2^3 = 8$ conditional expectations.}
\label{fig:lattice_shap_1}
\end{centering}
\end{figure}
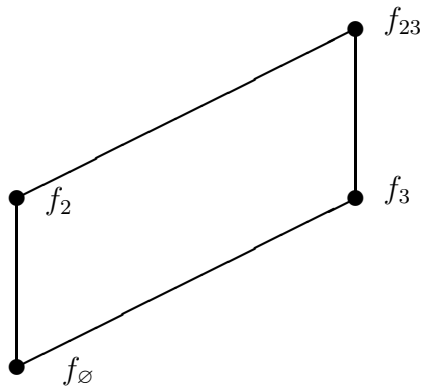
If $f(x) = f_{\varnothing} + f_2 + f_3 + f_{23}$, then the approximation is perfect so that 
$\E\left[\left(f(x) - \E\left[f(X_1, x_2, x_3) \right]\right)^2\right] = 0$. Using the notation and verbiage 
of \cite{hooker2004discovering}, we let $G_s(x)$ denote the approximation of $f(x)$ with $s$ and 
all of its supersets removed from the lattice and we refer to 
$\E\left[\left(f(x) - G_s(x)\right)^2\right]$ as the $\mathcal{L}_2$ cost of exclusion (L2COE) of set $s$. 
In this example, $G_s(x) = \E\left[f(X_1, x_2, x_3) \right]$ and L2COE$(s)$ = $\E\left[\left(f(x) - \E\left[f(X_1, x_2, x_3) \right]\right)^2\right]$.
We can use the L2COE as an interaction importance criteria in Algorithm \ref{alg:bfs}, replacing $\psi(s)$ = L2COE$(s)$. 
\cite{liu2006estimating} present a ``pick-freeze" algorithm for estimating L2COE. First, select a sample $Z$ of size $n \times p$. Then, select a 
sample $X$ of size $n \times \abs{s}$ where $\abs{s}$ is the size of the interaction term $s$. This gives an estimate
\begin{equation*}
\begin{aligned}
\textrm{L2COE} &= \frac{1}{n2^{\abs{s}}} \sum_{i = 1}^n \left( \sum_{r \subseteq s} f(x_i^r, z_i^{-r}) \right)^2
\end{aligned}
\end{equation*}
where $x_i^{r}$ refers to the $i$-th sample of $x$ for the features indexed by interaction $s$ and $z_i^{-r}$
refers to the $i$-th sample of $z$ for the features \textbf{not} indexed by interaction $s$.

To illustrate this algorithm, we implement a simple example where the active subsets in this case are $\{1\}$, $\{2\}$, $\{3\}$ and $\{2,3\}$.
\begin{equation*}
\begin{aligned}
X &= \left(X_1, X_2, X_3\right)\\
y &= f(X) = X_1 + X_2 + X_3 + X_2X_3\\
\end{aligned}
\end{equation*}

To test this algorithm, we conduct 100 simulations with $n=500$ using a custom R implementation of the algorithm. 
We approximate $V_t = \sum_{s \subseteq 2^{\Omega}} \hat{\sigma}_s(f_s)$ where $\hat{\sigma}_s(f_s) = \frac{1}{n}\sum_{i = 1}^n \hat{f}_s^2\left(x_i\right)$ 
and the expected values in $\hat{f}_s\left(x_i\right)$ are also calculated as empirical expectations with respect to the matrix $\mathbf{x} = \left(x_1, \cdots, x_n\right)'$. 
$X_2$ and $X_3$ have the same L2COE, and we observe that the correct ranking of either $\{\{2\}, \{3\}, \{1\}, \{2,3\}\}$ or $\{\{3\}, \{2\}, \{1\}, \{2,3\}\}$ in 
100\% of simulations.

Finally, with a list $\mathcal{S}$ of low-order interaction terms returned by the above algorithm, Shapley values can be estimated by computing the 
functional ANOVA terms for each of the subsets in $\mathcal{S}$ and combining them according to 
$\phi_i\left(f\right) = \sum_{j = 1}^p \frac{1}{j} \sum_{S \subseteq 2^{\Omega}: i \in S, \abs{S} = j} f_S$.

\subsubsection{Measures of effective dimensionality} \label{effective-dimension}

The above search procedure is helpful in identifying structure of the model, but it could be costly when the inductive bias is 
incorrect. \cite{saltelli2010variance} outline a sensitivity analysis technique that can be used to determine both factor 
sparsity as well as interaction order. \cite{kucherenko2009monte} introduce two definitions of ``effective dimensionality" which 
correspond roughly to the principles of effect sparsity and dominance by low-order interactions. 
The first measure, \textit{truncation} dimensionality, is defined as the value of $d_T$ such that 
\begin{equation*}
\begin{aligned}
\sum_{u \subseteq \{1,...,d_T\}} \sigma^2_u &\geq \left(1 - \varepsilon\right)\sigma^2
\end{aligned}
\end{equation*}
which corresponds broadly to the minimum number of factors required to explain $(1 - \varepsilon)$\% of the total variance. 
The second measure, \textit{superposition} dimensionality, is defined as the value of $d_S$ such that 
\begin{equation*}
\begin{aligned}
\sum_{0 < \abs{u} < d_S} \sigma^2_u &\geq \left(1 - \varepsilon\right)\sigma^2
\end{aligned}
\end{equation*}
which corresponds to the minimum interaction size required to explain $(1 - \varepsilon)$\% of the total variance. 

\cite{saltelli2010variance} show that two common variance-based sensitivity analysis measures can be used to 
assess $d_T$ and $d_S$. Common estimands for the first order and total variance-based sensitivities for feature $i$ are given by 
\begin{equation*}
\begin{aligned}
S_i &= \frac{\V_{X_i}\left(\E_{X_{-i}}\left(f(X) \mid X_i\right)\right)}{\V\left(f(X)\right)}\\
S_{T_i} &= \frac{\E_{X_{-i}}\left(\V_{X_i}\left(f(X) \mid X_{-i}\right)\right)}{\V\left(f(X)\right)} = 1 - \frac{\V_{X_{-i}}\left(\E_{X_{i}}\left(f(X) \mid X_{-i}\right)\right)}{\V\left(f(X)\right)}\\
\end{aligned}
\end{equation*}
There are a number of approaches to computing these indices using random or low-discrepancy samples for $X_i$ and $X_{-i}$ 
(see for example \cite{sobol2001global} and \cite{saltelli2002making}). The R package \texttt{sensitivity} (\cite{iooss2021sensitivity}) 
provides a convenient interface to compute both $S_i$ and $S_{T_i}$. Once these indices are computed, the importance of interactions 
can be determined by the magnitude of $S_i / S_{T_i}$ for each $i$ and the relative sparsity (or varying feature importance) can be determined 
by the distribution of $S_i$ or $S_{T_i}$. In addition to providing a computationally efficient gauge of sparsity and interaction importance, in some cases these 
indices could be used on their own to replace the numeric search algorithm in Section \ref{numeric-search}. For example, if 
$\sum_{i \in \{1,2,5\}} S_i \geq 1 - \varepsilon$, then the SHAP estimates can be computed with only 7 conditional expectations. 

\subsection{The impact of choosing a baseline distribution} \label{baseline-selection}

The SHAP estimand can be loosely characterized as ``the average effect of setting feature $i$ equal to its target value," which raises the 
question: average with respect to which distribution? In Section \ref{overview}, we referred to ``individual" and ``multiple" baselines as two different approaches to SHAP. However, we can unify these two approaches using the feature subset notation. In the Harsanyi dividend representation, each of the $f_u$ terms are a contrast of conditional expectations, which may be taken with respect to any distribution $p(X)$. In the ``multiple baseline" scenario commonly employed by SHAP users, the implied distribution $p(X)$ is the empirical distribution of features in the baseline set. Similarly, the ``single baseline" scenario constitutes a degenerate distribution for which $p(X) = \mathbf{1}\{X = x_{baseline}\}$. 

We examine the impact of the choice of baseline distribution by comparing Shapley values with $p = 3$ and $x_{target} = (1,1,1)$ 
with three different baseline distributions:
\begin{enumerate}[\qquad (a)]
\item \textbf{Global independent}: $p(X) \sim \mathcal{N}\left((0,0,0), I\right)$
\item \textbf{Global correlated}: $p(X) \sim \mathcal{N}\left((0,0,0), \begin{pmatrix} 1 & 0.9 & 0.5\\ 0.9 & 1 & 0.75 \\ 0.5 & 0.75 & 1 \end{pmatrix}\right)$
\item \textbf{Local independent}: $p(X) \sim \mathcal{N}\left(x_{target}, 0.25^2 I\right)$
\item \textbf{Single baseline}: $p(X_i) = \mathbf{1}\{X_i = x_{baseline, i}\}$
\end{enumerate}

We compute Shapley values for $x_{target}$ using the above three distributions on four functions presented in the table below.

\begin{table}
\begin{center}
\begin{tabular}{ |c|c|c|c|c|c| } 
\hline
$f(X)$ & Baseline type & $X_1$ & $X_2$ & $X_3$ \\
\hline
\multirow{4}{20em}{$-2X_1 + 1.5X_2 + 0.5X_3$} & A & -2.00 & 1.50 & 0.50 \\ 
& B & -0.39 & -0.03 & 0.41\\ 
& C & 0 & 0 & 0 \\
& D & -2.00 & 1.50 & 0.50 \\ 
\hline
\multirow{4}{20em}{$-2X_1 + 1.5X_2 + 0.5X_3 - 2X_2X_3$} & A & -2.00 & 0.50 & -0.50\\ 
& B & -0.47 & -0.01 & -0.02 \\ 
& C & 0 & 0 & 0 \\
& D & -2.00 & 0.50 & -0.50 \\ 
\hline
\multirow{4}{20em}{$-2\sin(X_1) + 1.5\abs{X_2} + 0.125X_3^2$} & A & -1.68 & 0.31 & -0.00 \\ 
& B & -0.78 & -0.48 & -0.13 \\ 
& C & -0.05 & 0.00 & -0.01 \\
& D & -1.68 & 1.50 & 0.12 \\ 
\hline
\multirow{4}{20em}{$-2\sin(X_1) + 1.5\abs{X_2} + 0.125X_3^2 + \cos(X_2X_3)$} & A & -1.69 & 0.22 & -0.09 \\ 
& B & -0.80 & -0.45 & -0.22 \\ 
& C & -0.05 & 0.02 & 0.01 \\
& D & -1.68 & 1.27 & -0.10 \\ 
\hline
\end{tabular}
\caption{Shapley values with different baseline distributions}
\label{shap-baselines}
\end{center}
\end{table}

The results (Table \ref{shap-baselines}) show that while each of these baselines could be viewed as reasonable or plausible in different circumstances, they are far from interchangeable. With strong correlation in the data, using the correlated conditional distribution to compute Shapley values has a profound effect on the estimated Shapley values relative to treating the features as independent. Similarly, a local baseline distribution centered around the target only estimates nonzero Shapley values for functions with strong evenness or nonlinearity. The single baseline estimates are perhaps the most intuitive to grasp, but they raise a crucial question of where to place the one representative baseline. Google's AI Explanations Whitepaper (\cite{google2020ai}) suggests the single baseline only when there is an ``informative reference" value for comparison.

\newpage

\section{Discussion} \label{discussion}

This work has explored the relationship between SHAP and the functional ANOVA, and design of experiments. A result of this connection is that decades of literature on computer experiments and function approximation may be brought to bear on questions of model interpretability. However, this connection also dashes any hope of settling the question of ``which Shapley values to use" due to the influence of the baseline distribution which is embedded in the estimand itself!

While SHAP is typically regarded as a tool for making modeling decisions ``interpretable," this interpretability is not free. As noted above, the estimand can be loosely messaged to stakeholders as ``the average effect of setting a feature equal to its target value," but this conceals decisions about the underlying distribution used in computing those averages. 

The connection between sensitivity analysis and model explainability is fascinating. We believe that computational methods used in sensitivity analysis (for example, quasi-monte carlo or low-discrepancy sampling) might be profitably applied to improve convergence of existing explainability methods.

An interesting future line of research would be to study the goals of model explainability and determine for which purposes SHAP is best suited. Modern model explainability tools are increasingly being used for legal or compliance purposes, such as algorithmic recourse (\cite{karimi2020survey}). Based on this review of SHAP, it is not immediately obvious that a series of contrasts of conditional expectations can provide the necessary information for algorithmic recourse. At the very least, this topic deserves further study, contrasting SHAP with methods such as counterfactual explanations (\cite{wachter2017counterfactual}).

\section{Acknowledgements}

This work was partially supported by NSF Grant DMS-1502640. We are grateful for helpful feedback provided by participants in a seminar at Arizona State University.
Code supporting the tables and computational examples in this paper is available at \url{https://github.com/andrewherren/shap-anova-examples}. The first author's use of Claude in a recent iteration of this manuscript surfaced the connection between functional ANOVA and the Harsanyi dividend.

\bibliography{SHAP_fANOVA}

\appendix \label{appendix}
\section{Derivation of the SHAP regression} \label{shap-reg-deriv}

Following \cite{rencher2008linear}, we can express the objective function as 
\begin{equation*}
\begin{aligned}
L(\beta, \lambda) &= \left(y - Z\beta\right)' W \left(y - Z\beta\right) + \lambda'\left(j'\beta - (y_t - y_b)\right)
\end{aligned}
\end{equation*}
where $\lambda$ is a Lagrange multiplier.

We can minimize this objective function by differentiating $L$ with respect to $\beta$ and $\lambda$, setting 
both partial derivatives equal to 0, and checking the determinant of the Hessian matrix. Solving for $\beta$ gives 
\begin{equation*}
\begin{aligned}
\hat{\beta} &= \left(Z'WZ\right)^{-1} \left( I - j A^{-1} j' \left(Z'WZ\right)^{-1} \right) Z'W y + \left(Z'WZ\right)^{-1} j A^{-1} \left(y_t - y_b\right)\\
A &= j' \left(Z'WZ\right)^{-1} j
\end{aligned}
\end{equation*}

We note, as do \cite{lundberg2017unified}, that $Z'WZ$ is a symmetric matrix of the form $a J + b I$, where 
$I$ is the $p$-dimension identity matrix and $J$ is a $p \times p$ matrix of all ones. Solving for $a$ and $b$, we get that
\begin{equation*}
\begin{aligned}
a &= \sum_{i=1}^{p-1} \frac{\left(i - 1\right)}{p\left(p - i\right)}\\
b &= \frac{p - 1}{p}
\end{aligned}
\end{equation*}

We can determine $\left(Z'WZ\right)^{-1}$ by setting $\left(Z'WZ\right)^{-1} = cJ + d I$ and solving
\begin{equation*}
\begin{aligned}
\left(Z'WZ\right) \left(Z'WZ\right)^{-1} &= I\\
\left(aJ + b I\right)\left(cJ + d I\right) & = acJ^2 + bcJ + ad J + bd I = pac J + bcJ + adJ + bdI = I\\
\end{aligned}
\end{equation*}

This implies that $pac + bc + ad = 0$ and $bd = 1$ so that $d = 1 / b = p / \left(p - 1\right)$ and $c = -\left(\frac{a}{b}\right)\left(\frac{1}{pa + b}\right)$.
Now observe that $j' \left(Z'WZ\right)^{-1} = (cp + d) j'$ and $j' \left(Z'WZ\right)^{-1} j = (cp + d)p$ and thus 
\begin{equation*}
\begin{aligned}
(j' \left(Z'WZ\right)^{-1} j)^{-1} &= \frac{1}{(cp + d)p}\\
j (j' \left(Z'WZ\right)^{-1} j)^{-1} j' \left(Z'WZ\right)^{-1} & = \frac{j j' \left(Z'WZ\right)^{-1}}{(cp + d)p} = J / p\\
\left(Z'WZ\right)^{-1} \left( I - j A^{-1} j' \left(Z'WZ\right)^{-1} \right) &= \left( cJ + d I \right) \left( I - \frac{1}{p} J \right)\\
& = d I  + c J - \frac{d}{p} J - c J = d I - \frac{d}{p} J = \frac{p}{p-1} I - \frac{1}{p - 1} J
\end{aligned}
\end{equation*}

Similarly, 
\begin{equation*}
\begin{aligned}
\left(Z'WZ\right)^{-1} j (j' \left(Z'WZ\right)^{-1} j)^{-1} & = \frac{\left(Z'WZ\right)^{-1} j'}{(cp + d)p} = j / p\\
\end{aligned}
\end{equation*}

Thus, the regression solution simplifies to 
\begin{equation*}
\begin{aligned}
\hat{\beta} &= \left(Z'WZ\right)^{-1} \left( I - j A^{-1} j' \left(Z'WZ\right)^{-1} \right) Z'W y + \left(Z'WZ\right)^{-1} j A^{-1} \left(y_t - y_b\right)\\
&= \left( \frac{p}{p-1} I - \frac{1}{p - 1} J \right) Z'W y + \frac{j\left(y_t - y_b\right)}{p}
\end{aligned}
\end{equation*}

$Z'W$ is a $p \times \left(2^p - 2\right)$ matrix in which the columns correspond to coalitions in the $Z$ matrix multiplied by the weight $w_i$ of that coalition. 
$\frac{p}{p-1} I - \frac{1}{p - 1} J$ is a square symmetric matrix. Letting $s$ represent the number of nonzero entries in a given column of $Z'W$, we see that 
the weights attached to those nonzero entires is $w(s) = \frac{\left(p-1\right)\left(p - s - 1\right)! \left(s - 1\right)!}{p!}$ so 
\begin{equation*}
\begin{aligned}
- \frac{1}{p - 1} J Z'W & = \begin{pmatrix} 
-\frac{s \left(p-1\right)\left(p - s - 1\right)! \left(s - 1\right)!}{\left(p-1\right) p!} & ... & -\frac{s \left(p-1\right)\left(p - s - 1\right)! \left(s - 1\right)!}{\left(p-1\right) p!} \\
... & ... & ... \\
-\frac{s \left(p-1\right)\left(p - s - 1\right)! \left(s - 1\right)!}{\left(p-1\right) p!} & ... & -\frac{s \left(p-1\right)\left(p - s - 1\right)! \left(s - 1\right)!}{\left(p-1\right) p!}
\end{pmatrix} = \begin{pmatrix} 
-\frac{s\left(p - s - 1\right)! \left(s - 1\right)!}{p!} & ... & -\frac{s\left(p - s - 1\right)! \left(s - 1\right)!}{p!} \\
... & ... & ... \\
-\frac{s\left(p - s - 1\right)! \left(s - 1\right)!}{p!} & ... & -\frac{s\left(p - s - 1\right)! \left(s - 1\right)!}{p!}
\end{pmatrix}
\end{aligned}
\end{equation*}
and 
\begin{equation*}
\begin{aligned}
\frac{p}{p - 1} I Z'W & = \begin{pmatrix} 
\frac{p \left(p-1\right)\left(p - s - 1\right)! \left(s - 1\right)!}{\left(p-1\right) p!} & ... & \frac{p \left(p-1\right)\left(p - s - 1\right)! \left(s - 1\right)!}{\left(p-1\right) p!} \\
... & ... & ... \\
\frac{p \left(p-1\right)\left(p - s - 1\right)! \left(s - 1\right)!}{\left(p-1\right) p!} & ... & \frac{p \left(p-1\right)\left(p - s - 1\right)! \left(s - 1\right)!}{\left(p-1\right) p!}
\end{pmatrix} = \begin{pmatrix} 
\frac{p \left(p - s - 1\right)! \left(s - 1\right)!}{p!} & ... & \frac{p\left(p - s - 1\right)! \left(s - 1\right)!}{p!} \\
... & ... & ... \\
\frac{p\left(p - s - 1\right)! \left(s - 1\right)!}{p!} & ... & \frac{p\left(p - s - 1\right)! \left(s - 1\right)!}{p!}
\end{pmatrix}
\end{aligned}
\end{equation*}

And thus the entries of matrix $B = \left( \frac{p}{p-1} I - \frac{1}{p - 1} J \right) Z'W$ are 
\begin{equation*}
\begin{aligned}
B_{ij} &= \begin{cases}
\frac{\left(p - s\right)! \left(s - 1\right)!}{p!} & Z_{ji} = 1\\
-\frac{s\left(p - s - 1\right)! \left(s - 1\right)!}{p!} = -\frac{\left(p - s - 1\right)! s!}{p!} & Z_{ji} = 0
\end{cases}
\end{aligned}
\end{equation*}
which correspond to Shapley weights with and without the feature of interest included in a coalition.
When $s = 0$ or $s = p - 1$, the Shapley weight is exactly $1/p$, which is the weight attached to 
$y_b$ and $y_t$ in the second term of regression coefficient solution. 
Thus, since $Z$ includes all $2^p - 2$ synthetic coalitions, the first term of $\hat{\beta}$ is a 
linear combination of all of the synthetic predictions $f(Z)$ and the second term of $\hat{\beta}$ 
is a column vector of $y_t / p - y_b / p$. 

Rather than view the regression weights as separate error and constraint terms, we can concatenate 
into one linear operation. Adding $y_b$ to the beginning of the $y$ vector and $y_t$ to the end, we have 
$y_{*} = \begin{pmatrix} y_b & y' & y_t \end{pmatrix}'$. Similarly, we can append the vector $j' / p$ to both ends 
of $B$, getting $B_{*} = \begin{pmatrix} -j' / p & B & j' / p \end{pmatrix}$. Then we see that 
$\hat{\beta} = B_{*} y_{*}$. Since each row of $B_{*}$ is a set of positive and negative Shapley weights and 
$y_{*}$ is the complete set of $2^p$ model predictions, we see that each entry $i$ in the 
$p$ rows of $\hat{\beta}$ correspond to the exact Shapley value, $\phi_i$, for feature $i$.

To see this illustrated, we return to the example from Figure \ref{fig:hypercube}. We can see that 
\begin{equation*}
\begin{aligned}
Z &= \begin{pmatrix} 
1 & 0 & 0 \\
0 & 1 & 0 \\
0 & 0 & 1 \\
1 & 1 & 0 \\
1 & 0 & 1 \\
0 & 1 & 1
\end{pmatrix} \; \; \; \; \; \;
W = \begin{pmatrix} 
1/3 & 0 & 0 & 0 & 0 & 0 \\
0 & 1/3 & 0 & 0 & 0 & 0 \\
0 & 0 & 1/3 & 0 & 0 & 0 \\
0 & 0 & 0 & 1/3 & 0 & 0 \\
0 & 0 & 0 & 0 & 1/3 & 0 \\
0 & 0 & 0 & 0 & 0 & 1/3
\end{pmatrix}
\end{aligned}
\end{equation*}
\begin{equation*}
\begin{aligned}
Z'WZ &= \begin{pmatrix} 
1 & 1/3 & 1/3 \\
1/3 & 1 & 1/3 \\
1/3 & 1/3 & 1
\end{pmatrix} \;\;\;\;\;
\left(Z'WZ\right)^{-1} = \begin{pmatrix} 
12/10 & -3/10 & -3/10 \\
-3/10 & 12/10 & -3/10 \\
-3/10 & -3/10 & 12/10
\end{pmatrix}\\
I - j A^{-1} j' \left(Z'WZ\right)^{-1} &= \begin{pmatrix} 
2/3 & -1/3 & -1/3 \\
-1/3 & 2/3 & -1/3 \\
-1/3 & -1/3 & 2/3
\end{pmatrix}\\
\left(Z'WZ\right)^{-1} \left( I - j A^{-1} j' \left(Z'WZ\right)^{-1} \right) &= 
\begin{pmatrix} 
1 & -1/2 & -1/2 \\
-1/2 & 1 & -1/2 \\
-1/2 & -1/2 & 1
\end{pmatrix}\\
\end{aligned}
\end{equation*}
\begin{equation*}
\begin{aligned}
Z'W &= 
\begin{pmatrix} 
1/3 & 0 & 0 & 1/3 & 1/3 & 0 \\
0 & 1/3 & 0 & 1/3 & 0 & 1/3 \\
0 & 0 & 1/3 & 0 & 1/3 & 1/3
\end{pmatrix}\\
\end{aligned}
\end{equation*}
\begin{equation*}
\begin{aligned}
\left(Z'WZ\right)^{-1} \left( I - j A^{-1} j' \left(Z'WZ\right)^{-1} \right) Z'W &= 
\begin{pmatrix} 
1/3 & -1/6 & -1/6 & 1/6 & 1/6 & -1/3\\
-1/6 & 1/3 & -1/6 & 1/6 & -1/3 & 1/6\\
-1/6 & -1/6 & 1/3 & -1/3 & 1/6 & 1/6
\end{pmatrix}\\
\left(Z'WZ\right)^{-1} j A^{-1} &= 
\begin{pmatrix} 
1/3 \\
1/3 \\
1/3
\end{pmatrix}\\
\end{aligned}
\end{equation*}
\begin{equation*}
\begin{aligned}
y &= \begin{pmatrix} f\left( t_1, b_2, b_3 \right) \\ f\left( b_1, t_2, b_3 \right) \\ f\left( b_1, b_2, t_3 \right) \\ f\left( t_1, t_2, b_3 \right) \\ f\left( t_1, b_2, t_3 \right) \\ f\left( b_1, t_2, t_3 \right) \end{pmatrix}\\
y_b &= f(b_1, b_2, b_3)\\
y_t &= f(t_1, t_2, t_3)
\end{aligned}
\end{equation*}

If we conduct the same concatenation as described above, we get 
\begin{equation*}
\begin{aligned}
y_{*} &= \begin{pmatrix} f\left( b_1, b_2, b_3 \right) \\ f\left( t_1, b_2, b_3 \right) \\ f\left( b_1, t_2, b_3 \right) \\ f\left( b_1, b_2, t_3 \right) \\ f\left( t_1, t_2, b_3 \right) \\ f\left( t_1, b_2, t_3 \right) \\ f\left( b_1, t_2, t_3 \right) \\ f\left( t_1, t_2, t_3 \right) \end{pmatrix}\\
B_{*} & = \begin{pmatrix} 
-1/3 & 1/3 & -1/6 & -1/6 & 1/6 & 1/6 & -1/3 & 1/3\\
-1/3 & -1/6 & 1/3 & -1/6 & 1/6 & -1/3 & 1/6 & 1/3\\
-1/3 & -1/6 & -1/6 & 1/3 & -1/3 & 1/6 & 1/6 & 1/3
\end{pmatrix}
\end{aligned}
\end{equation*}

And thus we find that
\begin{equation*}
\begin{aligned}
\phi &= \hat{\beta} = B_{*} y_{*}\\
\phi_1 &= \frac{1}{3} \left[ f\left( t_1, b_2, b_3 \right) - f\left( b_1, b_2, b_3 \right) \right] + 
\frac{1}{6} \left[ f\left( t_1, t_2, b_3 \right) - f\left( b_1, t_2, b_3 \right) \right] + \\
&\;\;\;\;\; \frac{1}{6} \left[ f\left( t_1, b_2, t_3 \right) - f\left( b_1, b_2, t_3 \right) \right] + 
\frac{1}{3} \left[ f\left( t_1, t_2, t_3 \right) - f\left( b_1, t_2, t_3 \right) \right]\\
\phi_2 &= \frac{1}{3} \left[ f\left( b_1, t_2, b_3 \right) - f\left( b_1, b_2, b_3 \right) \right] + 
\frac{1}{6} \left[ f\left( t_1, t_2, b_3 \right) - f\left( t_1, b_2, b_3 \right) \right] + \\
&\;\;\;\;\; \frac{1}{6} \left[ f\left( b_1, t_2, t_3 \right) - f\left( b_1, b_2, t_3 \right) \right] + 
\frac{1}{3} \left[ f\left( t_1, t_2, t_3 \right) - f\left( t_1, b_2, t_3 \right) \right]\\
\phi_3 &= \frac{1}{3} \left[ f\left( b_1, b_2, t_3 \right) - f\left( b_1, b_2, b_3 \right) \right] + 
\frac{1}{6} \left[ f\left( t_1, b_2, t_3 \right) - f\left( t_1, b_2, b_3 \right) \right] + \\
&\;\;\;\;\; \frac{1}{6} \left[ f\left( b_1, t_2, t_3 \right) - f\left( b_1, t_2, b_3 \right) \right] + 
\frac{1}{3} \left[ f\left( t_1, t_2, t_3 \right) - f\left( t_1, t_2, b_3 \right) \right]\\
\end{aligned}
\end{equation*}
and the Shapley values are the same as those calculated using the Shapley formula in Section \ref{shapley-value}. 
Thus, we see that the regression approximation yields exact Shapley values when the number of samples is 
exactly equal to $2^p$.

\end{document}

%% file: figure1.tex
\setlength{\unitlength}{2cm}
\begin{picture}(6,5)
\thicklines
\put(1,0.5){\circle*{0.15}}
\put(6,0.5){\circle*{0.15}}
\put(1,3.5){\circle*{0.15}}
\put(6,3.5){\circle*{0.15}}
\put(8,1.5){\circle*{0.15}}
\put(8,4.5){\circle*{0.15}}
\put(3,4.5){\circle*{0.15}}
\put(3,1.5){\circle*{0.15}}
\put(1,0.5){\line(5,0){5}}
\put(1,0.5){\line(0,3){3}}
\put(1,3.5){\line(5,0){5}}
\put(1,3.5){\line(2,1){2}}
\put(6,0.5){\line(2,1){2}}
\put(6,0.5){\line(0,3){3}}
\put(6,3.5){\line(2,1){2}}
\put(8,1.5){\line(0,3){3}}
\put(3,4.5){\line(5,0){5}}
\multiput(1,0.5)(0.3333333,0.1666667){6}{\line(2,1){0.2}}
\multiput(3,1.5)(0,0.3){10}{\line(0,3){0.2}}
\multiput(3,1.5)(0.5,0){10}{\line(5,0){0.2}}
\put(0.5,0.2){$f(b_1,b_2,b_3)$}
\put(5.6,0.2){$f(t_1,b_2,b_3)$}
\put(1.1,3.2){$f(b_1,t_2,b_3)$}
\put(2.85,1.2){$f(b_1,b_2,t_3)$}
\put(6.1,3.2){$f(t_1,t_2,b_3)$}
\put(8.1,1.4){$f(t_1,b_2,t_3)$}
\put(2.5,4.65){$f(b_1,t_2,t_3)$}
\put(8.1,4.4){$f(t_1,t_2,t_3)$}
\end{picture}

%% file: figure2.tex
\setlength{\unitlength}{2cm}
\begin{picture}(6,5)
\thicklines
\put(1,0.5){\circle*{0.15}}
\put(6,0.5){\circle*{0.15}}
\put(1,3.5){\circle*{0.15}}
\put(6,3.5){\circle*{0.15}}
\put(8,1.5){\circle*{0.15}}
\put(8,4.5){\circle*{0.15}}
\put(3,4.5){\circle*{0.15}}
\put(3,1.5){\circle*{0.15}}
\put(1,0.5){\line(5,0){5}}
\put(1,0.5){\line(0,3){3}}
\put(1,3.5){\line(5,0){5}}
\put(1,3.5){\line(2,1){2}}
\put(6,0.5){\line(2,1){2}}
\put(6,0.5){\line(0,3){3}}
\put(6,3.5){\line(2,1){2}}
\put(8,1.5){\line(0,3){3}}
\put(3,4.5){\line(5,0){5}}
\multiput(1,0.5)(0.3333333,0.1666667){6}{\line(2,1){0.2}}
\multiput(3,1.5)(0,0.3){10}{\line(0,3){0.2}}
\multiput(3,1.5)(0.5,0){10}{\line(5,0){0.2}}
\put(0.5,0.2){$\E\left[f(X_1,X_2,X_3)\right]$}
\put(5.6,0.2){$\E\left[f(t_1,X_2,X_3)\right]$}
\put(1.1,3.2){$\E\left[f(X_1,t_2,X_3)\right]$}
\put(2.85,1.2){$\E\left[f(X_1,X_2,t_3)\right]$}
\put(6.1,3.2){$\E\left[f(t_1,t_2,X_3)\right]$}
\put(8.1,1.4){$\E\left[f(t_1,X_2,t_3)\right]$}
\put(2.5,4.65){$\E\left[f(X_1,t_2,t_3)\right]$}
\put(8.1,4.4){$f(t_1,t_2,t_3)$}
\end{picture}

%% file: figure3.tex
\setlength{\unitlength}{1.5cm}
\begin{picture}(6,5)
\thicklines
\put(3,0.5){\circle*{0.15}}
\put(0,2){\circle*{0.15}}
\put(3,2){\circle*{0.15}}
\put(6,2){\circle*{0.15}}
\put(0,3.5){\circle*{0.15}}
\put(3,3.5){\circle*{0.15}}
\put(6,3.5){\circle*{0.15}}
\put(3,5){\circle*{0.15}}
\put(3,0.5){\line(-2,1){3}}
\put(3,0.5){\line(0,1){1.5}}
\put(3,0.5){\line(2,1){3}}
\put(0,2){\line(0,1){1.5}}
\put(0,2){\line(2,1){3}}
\put(3,2){\line(-2,1){3}}
\put(3,2){\line(2,1){3}}
\put(6,2){\line(-2,1){3}}
\put(6,2){\line(0,1){1.5}}
\put(0,3.5){\line(2,1){3}}
\put(3,3.5){\line(0,1){1.5}}
\put(6,3.5){\line(-2,1){3}}
\put(3.4,0.4){$f_{\varnothing}$}
\put(0.25,1.9){$f_1$}
\put(3.25,1.9){$f_2$}
\put(6.25,2){$f_3$}
\put(0.25,3.4){$f_{12}$}
\put(3.25,3.4){$f_{13}$}
\put(6.25,3.5){$f_{23}$}
\put(3.2, 5){$f_{123}$}
\end{picture}

%% file: figure4.tex
\setlength{\unitlength}{1.5cm}
\begin{picture}(6,4)
\thicklines
\put(3,0.5){\circle*{0.15}}
\put(3,2){\circle*{0.15}}
\put(6,2){\circle*{0.15}}
\put(6,3.5){\circle*{0.15}}
\put(3,0.5){\line(0,1){1.5}}
\put(3,0.5){\line(2,1){3}}
\put(3,2){\line(2,1){3}}
\put(6,2){\line(0,1){1.5}}
\put(3.4,0.4){$f_{\varnothing}$}
\put(3.25,1.9){$f_2$}
\put(6.25,2){$f_3$}
\put(6.25,3.5){$f_{23}$}
\end{picture}